\documentstyle[amssymb,epsf]{article}

\begin{document}

\title{Anomalous scattering rate and microwave absorption  in $%
Bi_{2}Sr_{2}CaCu_{2}O_{8+\delta }$ and $YBa_{2}Cu_{3}O_{7-\delta }$}
\author{C.Kusko$^{1,2}$, Z.Zhai$^{1}$, R.S.Markiewicz$^{1,2}$  
and S.Sridhar$^{1}$\\
\\
{\it $^1$ Physics Department, Northeastern University, Boston MA02115}\\
{\it $^2$ Barnett Institute, Northeastern University, Boston MA02115}}
\maketitle
\begin{abstract}
We determine the scattering rate from microwave measurements for an
optimally doped Bi-2212 single crystal, using a simple two fluid model with
a d-wave symmetry order parameter. In the superconducting state, the
calculated scattering rate is three orders of magnitude smaller than that
determined from ARPES experiments. A similar anomalously large decrease in
the scattering rate is also required to explain the data within a
gap-quasiparticle scenario for other HTS, such as $YBa_{2}Cu_{3}O_{7-\delta
} $. The results suggest that the assumption of normal excitations vanishing
at low $T$ is invalid and an additional charge mode is responsible for the
microwave absorption.
\end{abstract}

One of the outstanding problems of the high $T_{c}$ cuprates is to
understand the mechanism of the linear microwave properties. Whereas there
were attempts to explain the microwave loss using a quasiparticle
description with an order parameter satisfying a d-wave (or mixed $d+s$)
symmetry, recent analyses of microwave \cite{Sridhar00} and terahertz
Time-Domain Spectroscopy (THz TDS) \cite{Corson00} clearly indicate that an
additional non-quasiparticle collective mode is responsible for the large
electromagnetic absorption at low temperatures.

In this paper we analyze the temperature dependence of the complex
conductivity $\tilde{\sigma}=\sigma _{1}+i\sigma _{2}$ \cite{MattisBardeen58}
for optimally doped high quality $Bi:2212$ and YBCO single crystals. In $%
Bi:2212$, while the behavior of $\sigma _{2}$ is consistent with a $%
d_{x^{2}-y^{2}}$ gap \cite{Jacobs95}, the nearly linear increase of the
normal conductivity $\sigma _{1}$ to a large value at low temperatures is in
direct contradiction with the conventional picture of the competition \cite
{Bonn} between the decrease in the number of excited quasiparticles and the dramatic
increase of their scattering rate.

Above $T_{c}$, assuming a local $j-\hat{E}$ relation in the skin depth
limit, the normal state surface resistance can be written as $R_{n}=\sqrt{%
\omega \mu _{0}\rho _{n}/2}$, where the microwave normal state resistivity
is expected to be the same as the DC resistivity in the so-called
Hagens-Rubens limit, and hence $\rho _{n}=2\Gamma \mu _{0}\lambda (0)^{2}$.
Here $\lambda (0)$ is the {\em zero} temperature penetration depth, and $%
\Gamma $ is the normal state scattering rate. Typically the normal
resistivity $\rho _{n}$ obeys linear temperature dependence $\rho _{n}=\rho
_{0}+\gamma T$. Therefore the resistivity values can be translated into the
inelastic scattering rates given by $\Gamma =\frac{\gamma T}{2\mu
_{0}\lambda ^{2}(0)}$, where $\gamma $ is the coefficient of the linear term
in $\rho _{n}$ {\em vs.} $T$.

Below $T_{c}$, the scattering rate can be obtained from the complex
conductivity $\tilde{\sigma}=\sigma _{1}+i\sigma _{2}$. In a
phenomenological ``two-fluid'' model the high frequency conductivity $\tilde{%
\sigma}(\omega ,T)$ can be expressed as:

\begin{equation}
\tilde{\sigma}(\omega ,T)=\sigma _{1}+i\sigma _{2}=\frac{ne^{2}}{m}\left[ 
\frac{n_{n}(T)}{i\omega +1/\tau (T)}+\frac{in_{s}(T)}{\omega }\right]
\label{sigma}
\end{equation}

\noindent where $n_{n}$ and $n_{s}$ represent the fractions of normal and
superconducting quasiparticles (with $n_{n}+n_{s}=1$), and $\tau $ is the
relaxation time for the normal electrons. In this model, the normal
electrons have damping with the usual Drude conductivity at high
frequencies, and the superconducting electrons have inertia but no damping.
In the local London limit, defined by the condition $\xi \ll \ell \ll
\lambda $ which is well satisfied by $YBCO$, $n_{n}$ and $n_{s}$ can be
calculated using the Mattis-Bardeen expression \cite{MattisBardeen58} :

\begin{equation}
n_{s}=(1-n_{n})=1-2\left\langle \int_{0}^{\infty }\left( -\,\frac{\partial f%
}{\partial E}\right) d\epsilon \right\rangle  \label{gap}
\end{equation}

where the quasiparticle energy is $E=\sqrt{\epsilon ^{2}+\Delta _{k}^{2}}$
and $\Delta _{k}$ is the superconducting order parameter with a d-wave
symmetry. We assume a $BCS$ temperature dependence for the $d_{x^{2}-y^{2}}$
gap parameter $\Delta (T,\phi )=\Delta _{d}(T)\cos (2\phi )$. The $d$-wave
model describes well the low $T$ behavior of $\lambda (T) \propto T$ for $%
Bi_{2}Sr_{2}CaCu_{2}O_{8+\delta }$ ($Bi:2212$)\cite{Jacobs95}.

From Eq. \ref{sigma} we can calculate the inverse lifetime in the
superconducting state using the approximation that the frequency $\omega$ is
negligible in comparison with the scattering rate.

\begin{figure}
\leavevmode
   \epsfxsize=0.8\textwidth\epsfbox{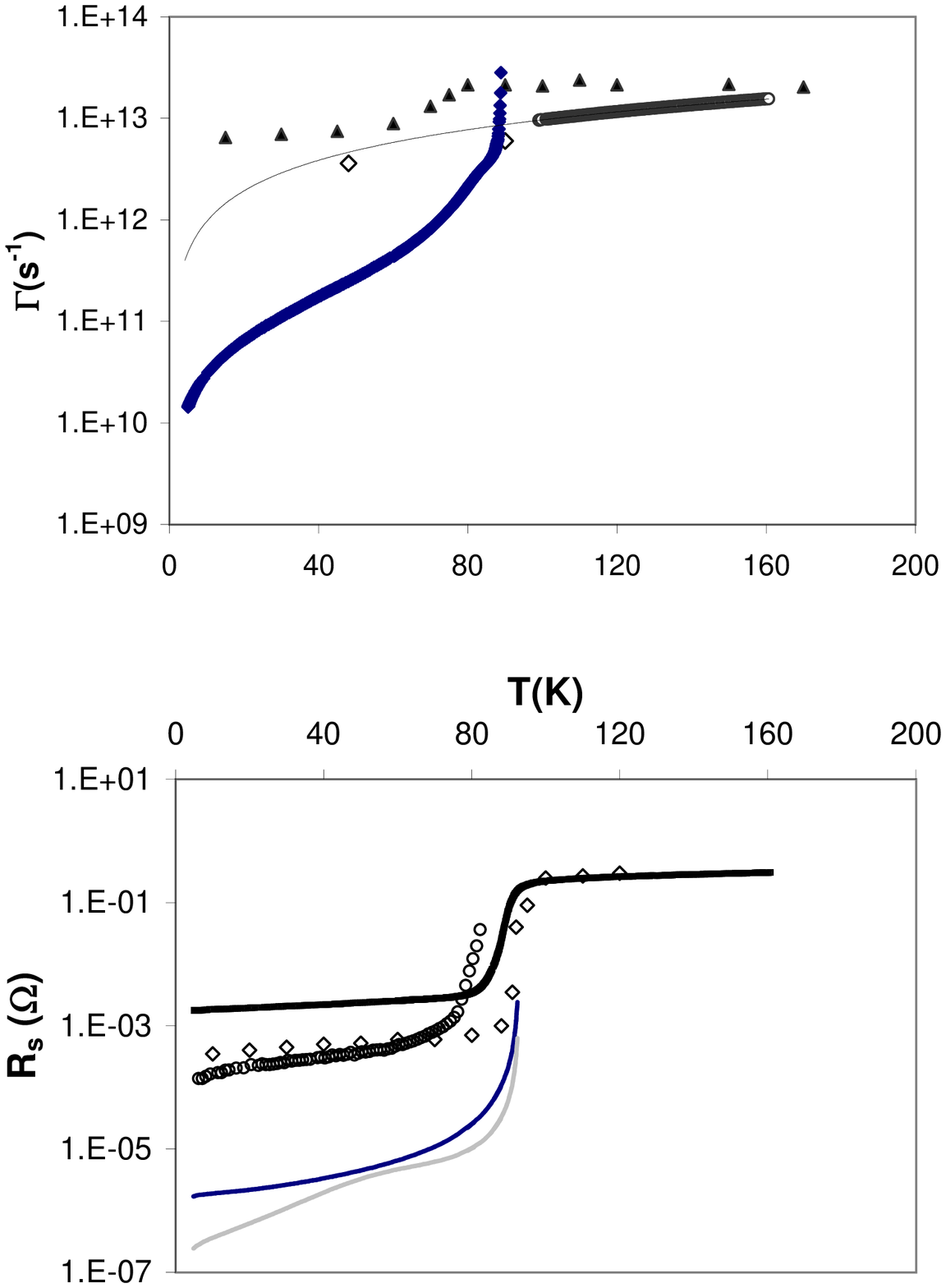} 
\caption{Upper panel: The calculated scattering rate for BSCCO using a two fluid
model and a d-wave order parameter (filled diamonds), the experimental 
scattering rate obtained from ARPES at $(\pi,0)$ points in the Brillouin zone 
(filled triangles) and the one measured at the nodal points $(\pi/2,\pi/2)$ (open diamonds).
The thin line represents the extrapolation of the normal state scattering rate (grey thick 
line) below $T_c$ according to the marginal Fermi liquid phenomenology.
Lower panel: 
The experimental surface resistance determined from our microwave measurements 
(black thick line). The open diamonds are the $R_s$ from [9] and the open circles are the $R_s$ from [10]. 
The black thin line represents
the calculated surface resistance
using the scattering rate obtained from the nodal points $(\pi/2,\pi/2)$
 ARPES measurements,
and the grey thin line is the one using the scattering rate from the $(\pi,0)$
ARPES measurements.}
\label{bsccogama}
\end{figure}

In Fig. \ref{bsccogama} we show the scattering rate $\Gamma (T)$ for $%
Bi:2212 $ calculated using a d-wave symmetry model . $\Gamma (T<T_{c})$
drops rapidly at $T_{c}$ but does not reach a limiting value, and instead
continues to decrease rapidly at lower temperatures. This very large
variation of $\Gamma $ is necessary to quantitatively describe the large
values of the microwave absorption within the quasiparticle framework.

Recently the scattering rate in $Bi:2212$ has been measured by angle
resolved photo-emission techniques (ARPES) \cite{Norman98,Valla99}. The
resulting experimentally measured values of the scattering rate $\Gamma $ by
ARPES are also shown in Fig.\ref{bsccogama}. It is more appropriate to
compare the microwave data with the ARPES results of Valla, et al.\cite
{Valla99}. The ARPES results of \cite{Norman98} are measured at the points $%
(\pi ,0)$ in the Brillouin zone where the d-wave superconducting gap has the
largest value, while the results of \cite{Valla99} are measured at the nodal 
$(\pi/2 ,\pi/2)$ points of the superconducting gap which gives the dominant
contribution in the transport measurements. The microwave measurements,
which are integrated over the entire FS are dominated, particularly in the
superconducting state, by the nodal quasiparticles.

Valla et al. \cite{Valla99} find that in the normal state the energy and
temperature dependence of the scattering rate are consistent with a Marginal
Fermi Liquid phenomenology $1/\tau =\max (\omega ,T)$ and this is not
affected by the superconducting transition. The normal state values measured
by both the microwave and ARPES\ techniques are in good agreement. However,
below $T_{c}$ the ARPES scattering rate continues to drop but only linearly
with $T$. The strong temperature dependence of $\Gamma $ ensures that these
measurements are not resolution limited, even at the lowest temperatures.
Clearly the calculated microwave $\Gamma $ and the measured ARPES\ $\Gamma $
are significantly different at $T<T_{c}$ , although they agree well in the
normal state.

The ARPES results of \cite{Norman98}, at $(\pi ,0)$, are considerably
broadened, and apparently resolution limited at the lowest temperatures.
(The Gaussian lineshape reported by Ding, et al.\cite{DEW} suggests
inhomogeneous broadening, with the intrinsic linewidth considerably
smaller.) Nevertheless ARPES is unlikely to achieve the resolution ($<0.1meV$%
) required to measure the extremely low scattering rates implied by the
microwave measurements.

It is worth pointing out that using the measured ARPES $\Gamma $ would lead
to a microwave absorption (represented by the surface resistance $R_{s}$)
orders of magnitude less than measured. This is shown in Fig.\ref{bsccogama}
where we have calculated the $R_{s}\,$using the scattering rates obtained
from both sets of ARPES measurements from \cite{Norman98} and \cite{Valla99}
and $\lambda (0)=2600\AA $. The resulting calculated $R_{s}$ is clearly
several orders of magnitude lower than the measured data.

Lee, et. al., \cite{Lee96} have also measured $R_{s}$ and $\lambda $ of $%
Bi:2212$ crystals at 14.4, 24.6 and 34.7 GHz. At $14.4$ $GHz$ their $R_{s}$
values are lower than ours, although the $\lambda (T)\,$data are essentially
the same. Their data are also included in Fig.\ref{bsccogama} and show that
the conclusions arrived at are still unchanged, i.e. the calculated values
using the above scattering rates are several orders of magnitude lower than
the measured ones.

\begin{figure}
\leavevmode
   \epsfxsize=0.8\textwidth\epsfbox{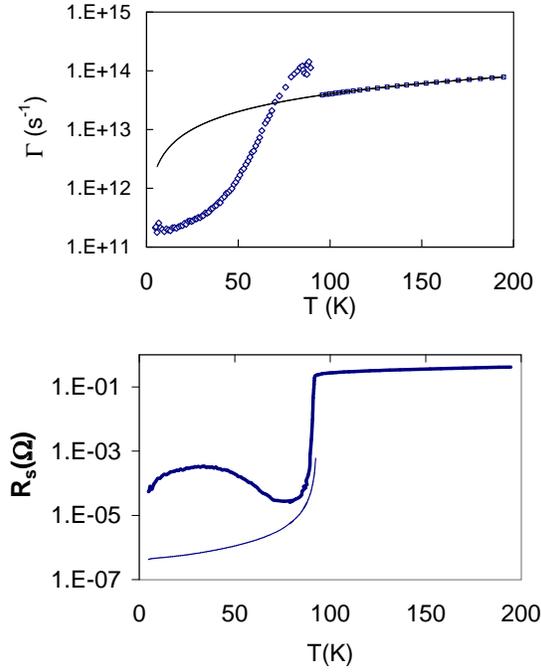} 
    \caption{Upper panel: The calculated scattering rate for YBCO determined from microwave 
measurements using 
a two fluid model and d-wave order parameter. The open diamonds represent the scattering rate
in the superconducting state, the filled squares are the scattering rates in the normal state.
The thin solid  line is the linear extrapolation of the normal state scattering rate below $T_c$.
Lower panel: The experimental surface resistance determined from our microwave measurements 
(black thick line). The thin solid line is the calculated surface resistance using the linear 
extrapolation of the normal state scattering data.}
\label{ybcogama}
\end{figure}

While we have specifically discussed the case of $Bi:2212$, similar
conclusions also hold for other superconductors. Fig. \ref{ybcogama} shows
the d-wave calculated $\Gamma (T)$ for $Y:123$. Here again the data require
that the scattering rate drop by $10^{2}$ from the normal state value.
However, there are no ARPES determinations of the scattering rate to compare
with. One does not expect that the scattering rate would differ by orders of
magnitude between YBCO and BSCCO. We note that the situation is
substantially more complicated in YBCO than in BSCCO, due to the presence of
multiple conductivity peaks \cite{Srikanth98} leading to a picture not
consistent with a single d-wave superconducting order parameter. A model 
based on a nested Fermi surface with a d-wave superconducting order parameter
has been developed in order to explain the anomalous behavior of the
microwave conductivity and surface resistance in YBCO \cite{Rie99}

Clearly then a quasiparticle mechanism, with its attendant assumption that $%
n_{n}$ decrease with decreasing $T$, is violated in $Bi:2212$ and other high
temperature superconductors. This strongly indicates that the microwave
absorption mechanism is distinctly different from a gap-quasiparticle
scenario. A number of alternative mechanisms are suggested. A possible
alternative is a collective mode occuring in the superconducting state.
Indeed collective modes such as density waves are extremely likely in low
dimensional materials. The contribution of a collective excitation due to an
overdamped charge density wave (CDW)-like mode turning on at the
superconducting transition has been already employed to describe this
anomalous temperature dependence of $\sigma _{1}$\cite{Sridhar00} in $%
Bi:2212 $. This is described in terms of an additional CDW-like contribution
with a phenomenological temperature dependent dielectric constant $\epsilon
(T)=\epsilon (0)(1-t^{2})$, where $t=T/T_{c}$ and $\epsilon (0)\approx
10^{8} $ which accounts for the linear increase of the $\sigma _{1}$ in the
superconductive state. Such a large value of dielectric constant is not
uncommon in low-dimensional CDW systems\cite{Gruner}.

It should be noted that dielectric modes have been recently observed by us
at microwave frequencies in non-superconducting $YBa_{2}Cu_{3}O_{6.0}\,$\cite
{Zhai-ybco60}, and at THz frequencies in superconducting $%
YBa_{2}Cu_{3}O_{6.95}\,$films by Wilke, et. al., \cite{Wilke00}. Possible
origins for dielectric modes in high-$T_{c}$ cuprates have been earlier
proposed in the literature \cite{Shenoy}.

At terahertz frequencies Corson et al.\cite{Corson00} have found a
significant excess non-quasiparticle contribution to $\sigma _{1}$ which
represents 30\% of the condensate spectral weight. They calculated the
difference $\sigma _{cm}$ between the measured $\sigma _{1}$ and the
conductivity due to quasiparticles calculated assuming a linear temperature
dependence of the inverse lifetime. They have ascribed this difference to a
collective mode which could be due to either classical phase fluctuations of
the superconducting condensate \cite{Stroud00} or to Josephson coupling
associated with spatial variation of the superfluid density \cite{Marel96}.
Altough at terahertz frequencies the quasiparticle contribution is
important, whereas in the microwave regime this is negligible, it is worth
pointing out the similarity between the THz $\sigma _{cm}$ and the microwave 
$\sigma _{1}$. This strongly suggests that the charge mode which is the
dominant mechanism for electromagnetic absorption at microwaves frequencies
as proposed in ref.\cite{Zhai00} is also present at much higher THz
frequencies. 

The presence of the nanoscale phase separation (stripes) in the high-$T_{c}$
superconductors naturally leads to an inhomogeneous ground state, and could
lead to a mechanism explaining the anomalous non-quasiparticle absorption
and finite normal conductivity at low $T.$ 

The huge discrepancy between the calculated normal microwave conductivity
using a conventional quasiparticle scenario and ARPES scattering rates and
the experimental one clearly indicates the breakdown of the quasiparticle
description of the high-$T_{c}$ superconductors at microwave frequencies. A
plausible candidate for the observed microwave absorption is a charge mode,
which may also be responsible for the electromagnetic response over a much
wider frequency range.

This work was supported by ONR N00014-00-1-0002 and NSF-9711910.

\end{document}